\documentclass[showkeys,superscriptaddress,amsmath,amssymb,aps,prx,twocolumn]{revtex4-1}

\usepackage[utf8x]{inputenc}
\usepackage{graphicx}% Include figure files
\usepackage{dcolumn}% Align table columns on decimal point
\usepackage{bm}% bold math
\usepackage{xcolor}
\usepackage{amsmath}
\usepackage{chngcntr}
\usepackage{nicefrac}
\usepackage[colorlinks,linkcolor=blue,hyperindex,CJKbookmarks]{hyperref}

\begin{document}

\title{Dispersionless orbital excitations in (Li,Fe)OHFeSe superconductors}

\author{Qian Xiao}\thanks{These authors contributed equally to this work.}
\affiliation{International Center for Quantum Materials, School of Physics, Peking University, Beijing 100871, China}

\author{Wenliang Zhang}\thanks{These authors contributed equally to this work.}
\author{Teguh Citra Asmara}\thanks{These authors contributed equally to this work.}
\affiliation{Photon Science Division, Swiss Light Source, Paul Scherrer Institut, CH-5232 Villigen PSI, Switzerland}

\author{Dong Li}\thanks{These authors contributed equally to this work.}
\affiliation{Beijing National Laboratory for Condensed Matter Physics, Institute of Physics, Chinese Academy of Sciences, Beijing 100190, China}
\affiliation{School of Physical Sciences, University of Chinese Academy of Sciences, Beijing 100049, China}

\author{Qizhi Li}
\author{Shilong Zhang}
\affiliation{International Center for Quantum Materials, School of Physics, Peking University, Beijing 100871, China}

\author{Yi Tseng}
\affiliation{Photon Science Division, Swiss Light Source, Paul Scherrer Institut, CH-5232 Villigen PSI, Switzerland}

\author{Xiaoli Dong}
\affiliation{Beijing National Laboratory for Condensed Matter Physics, Institute of Physics, Chinese Academy of Sciences, Beijing 100190, China}
\affiliation{School of Physical Sciences, University of Chinese Academy of Sciences, Beijing 100049, China}
\affiliation{Songshan Lake Materials Laboratory, Dongguan, Guangdong 523808, China}

\author{Yao Wang}
\affiliation{Department of Physics and Astronomy, Clemson University, Clemson, South Carolina 29631, USA}

\author{Cheng-Chien Chen}
\affiliation{Department of Physics, University of Alabama at Birmingham, Birmingham, Alabama 35294, USA}

\author{Thorsten Schmitt}
\email{thorsten.schmitt@psi.ch}
\affiliation{Photon Science Division, Swiss Light Source, Paul Scherrer Institut, CH-5232 Villigen PSI, Switzerland}

\author{Yingying Peng}
\email{yingying.peng@pku.edu.cn}
\affiliation{International Center for Quantum Materials, School of Physics, Peking University, Beijing 100871, China}

\date{\today}

\begin{abstract}
The superconducting critical temperature $T_{\mathrm{c}}$ of intercalated iron-selenide superconductor (Li,Fe)OHFeSe (FeSe11111) can be increased to 42\,K from 8\,K of bulk FeSe. It shows remarkably similar electronic properties as the high-$T_{\mathrm{c}}$ monolayer FeSe and provides a bulk counterpart to investigate the origin of enhanced superconductivity. Unraveling the nature of excitations is crucial for understanding the pairing mechanism in high-$T_{\mathrm{c}}$ iron selenides. Here we use resonant inelastic x-ray scattering (RIXS) to investigate the excitations in FeSe11111. Our high-quality data exhibit several Raman-like excitations, which are dispersionless and isotropic in momentum transfer and robust against varying $T_{\mathrm{c}}$. Using atomic multiplet calculations, we assign the low-energy $\sim 0.3$ and 0.7\,eV Raman peaks as local $e_g-e_g$ and $e_g-t_{2g}$ orbital excitations. The intensity of these two features decreases with increasing temperature, suggesting a primary contribution of the orbital fluctuations. 
Our results highlight the importance of orbital degree of freedom for high-$T_{\mathrm{c}}$ iron selenides.

\end{abstract}

\maketitle

\noindent
{\bf {INTRODUCTION}}\\

After more than a decade of the discovery of Fe-based superconductors \cite{HosonoFeSC}, 
the origin of superconductivity is still under debate \cite{wang2011electron}. Spin fluctuations, between the hole pockets at the Brillouin zone center and the electron pockets at the zone edges, have been considered as a candidate medium of the electron pairing, which typically leads to sign-reversed s$\pm$ pairing\,\cite{PhysRevLett.Mazin.SCpairing,Hirschfeld_2011}.
This mechanism has been supported by the spin resonance mode observed by neutron scattering measurements below the superconducting gap\,\cite{2008.nat.INS.BKFA}.
Apart from the important role of magnetism\,\cite{P.C.Dai.RMP2015}, other factors and mechanisms have also been proposed in Fe-based superconductors\,\cite{PhysRevB_orbitalfluctuation1,PhysRevB_orbitalfluctuation2,PhysRevLett_orbitalfluctuation,H.Gretasson2011prb,chu2012sci,ARPES.2015.MYi}. For example, orbital fluctuations have been proposed to enhance spin fluctuation-mediated pairing\,\cite{PhysRevB_orbitalfluctuation2} or directly account for the high $T_{\mathrm{c}}$ in iron pnictides \,\cite{PhysRevB_orbitalfluctuation1,PhysRevLett_orbitalfluctuation}. Angle-resolved photoemission spectroscopy (ARPES) measurements on iron chalcogenides revealed the disappearance of $d_{xy}$-orbital spectral weight with increasing temperature, implying universal orbital-selective correlation effects\,\cite{ARPES.2015.MYi}. 

Among the Fe-based superconductors, FeSe has the simplest crystal structure and provides an ideal platform for the study of the pairing mechanism\,\cite{hsu2008superconductivity}. While the $T_{\mathrm{c}}$ of bulk FeSe is only $\sim$ 8\,K, it can be enhanced by inter-layer interactions: both the intercalation between FeSe layers and ${\mathrm{SrTiO}}_{3}$ (STO) substrates for the monolayer FeSe can enhance the $T_{\mathrm{c}}$ by a factor of over four\,\cite{WangQingYan.2012cpl.monolayer.2gap,RevModPhysFeSe,lu2015,Dong.prb}. Two possible pairing mechanisms were proposed for the enhanced superconductivity in monolayer FeSe -- interfacial charge-transfer and electron-phonon coupling (EPC)\,\cite{WangQingYan.2012cpl.monolayer.2gap,2014.nat.monolayerFeSe.ARPES,Rademaker_2016,2017.prl.monolyerFeSe.ARPES.,zhao2018direct_SA,peng2020picoscale_SA}. However, these scenarios call for verification in an effective monolayer material without the substrate and related phonons.

LiOH-intercalated FeSe (FeSe11111) is a pure bulk single-crystalline superconductor with a $T_{\mathrm{c}}$ of over 40\,K\,\cite{lu2015,Dong.prb}, which satisfies these conditions. 
Due to the intercalation, the distance between FeSe layers becomes as large as 9\,\AA\, in FeSe11111, leading to a highly two-dimensional structure. 
ARPES experiments also reflect the similarity between FeSe11111 and monolayer FeSe/STO in terms of Fermi surface topology, band structure and the gap symmetry\,\cite{zhao2016.natcom}. 
As such, FeSe11111 resembles the monolayer FeSe, avoiding the interface to the STO substrate, and therefore provides a unique opportunity to elucidate the origin of the increased $T_c$ evolving from bulk FeSe to monolayer FeSe/STO. Studies of FeSe11111 recently lead to an even more exciting observation of a Majorana zero mode implying its nontrivial topology\,\cite{majorana2018FeSe11111}. Therefore, it is appealing to investigate the excitations in FeSe11111 as a crucial step towards understanding the pairing mechanism in iron chalcogenides.

\begin{figure*}[htbp]
\centering
\includegraphics[width=\linewidth]{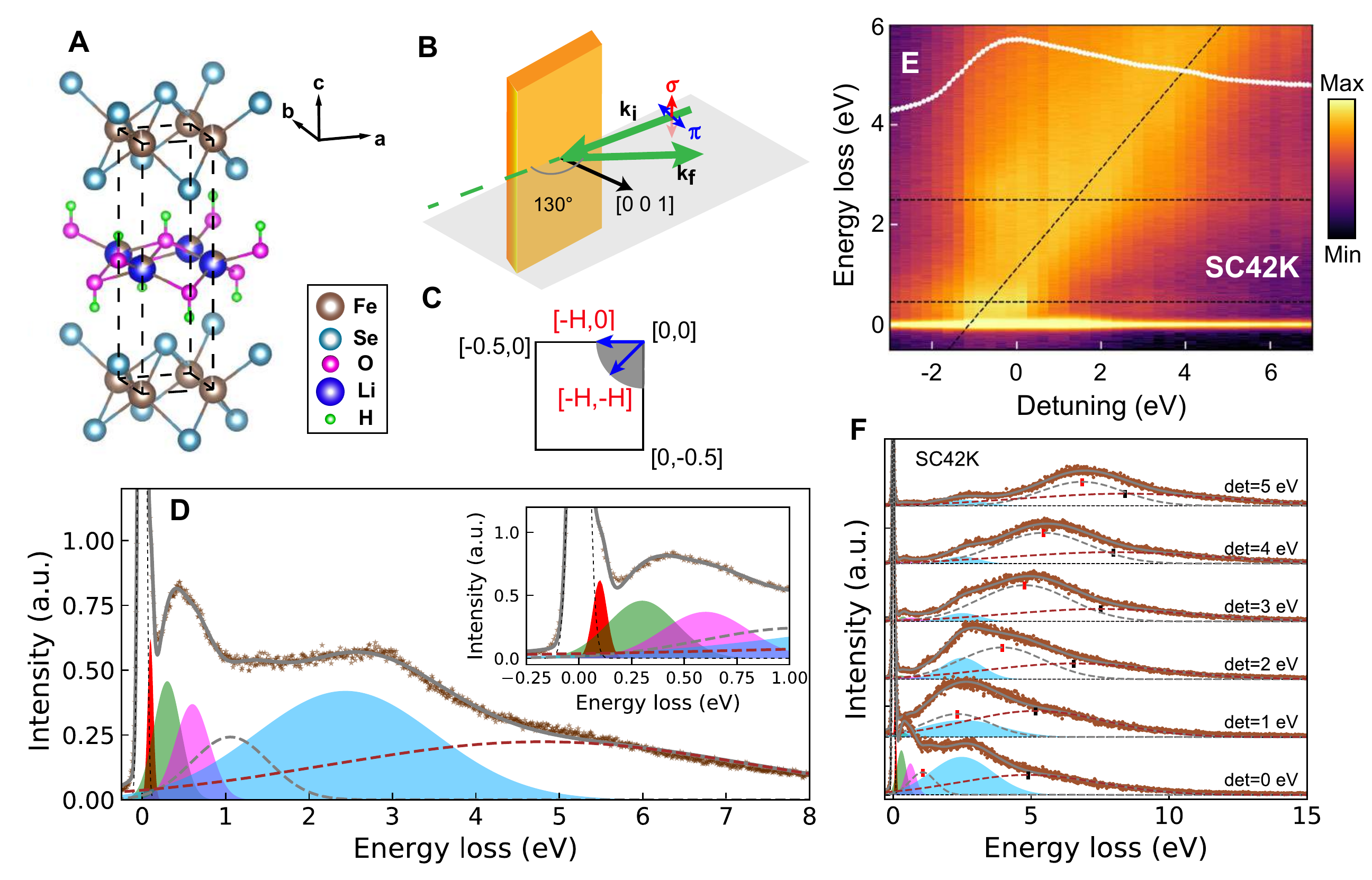}
\caption{{\bf RIXS spectra of FeSe11111.} ({\bf A}) Crystal structure of FeSe11111 in the one-Fe unit cell notation\,\cite{VESTA}. ({\bf B})  RIXS experimental geometry. The sample a-b plane lies perpendicular to the scattering plane. ({\bf C}) Reciprocal space in the one-Fe unit cell representation. The gray shaded area indicates the accessible momenta at Fe $L_3$-edge. ({\bf D}) A representative RIXS spectrum of SC28K taken at a photon energy of 709\,eV at 50\,K. The fitting components are overlaid and the gray solid line is the fit to data. (Inset) An enlarged view of the spectra below 1\,eV. ({\bf E}) RIXS intensity map (in logscale) versus energy loss and detuning energy across the Fe $L_3$-edge at 25\,K for SC42K. The white line is Fe $L_3$-edge X-ray absorption spectrum, measured via total fluorescence yield (TFY). The black dashed lines are guides to the eyes. ({\bf F}) Selected RIXS spectra of incident-energy detuning measurements. Each spectrum and fitting components were shifted vertically for clarity. Vertical ticks indicate two fluorescence-like features. 
\label{fig1}}
\end{figure*}

To identify the nature of excitations in FeSe11111 and their connection to superconductivity, we employ the resonant inelastic x-ray scattering (RIXS) technique, which is sensitive to multiple excitations\,\cite{RevModPhys.RIXS.2011} and has been widely used to study Fe-based superconductors\,\cite{yang2009prb,BFA.RIXS.natcom,M.C.Rahn.prb.FeSe,Jonathan.BFAP,Jonathan2021.natcom}.
We uncover the excitations in FeSe11111 films grown on a LaAlO$_3$ substrate with $T_{\mathrm{c}} \simeq$ 42\,K (SC42K) and $T_c \simeq$ 28\,K (SC28K) (``MATERIALS AND METHODS''). The higher $T_{\mathrm{c}}$ accompanies a larger $c$ lattice parameter and a lower Fe vacancy concentration in the FeSe-layer\,\cite{FeSe11111Tc42K,Yiyuan.CPL,FeVacancy2019prm}. We identified six excitations in both samples, including four Raman-like excitations and two fluorescence-like excitations. The four Raman features are located at $\sim$ 0.1\,eV, $\sim$ 0.3\,eV, $\sim$ 0.7\,eV, and $\sim$ 2.5\,eV. Through their momentum and temperature dependence, we attribute the $\sim$ 0.3\,eV and $\sim$ 0.7\,eV features to orbital excitations, and the $\sim$ 2.5\,eV feature to multiplet spin excitations. These orbital excitations are sensitive to the distorted tetrahedral crystal field and, accordingly, the lattice geometry. Therefore, our study suggests the lattice-induced orbital fluctuations as an ingredient to enhance the $T_c$ of iron chalcogenides.\\

\begin{figure*}[htbp]
\centering
\includegraphics[width=\linewidth]{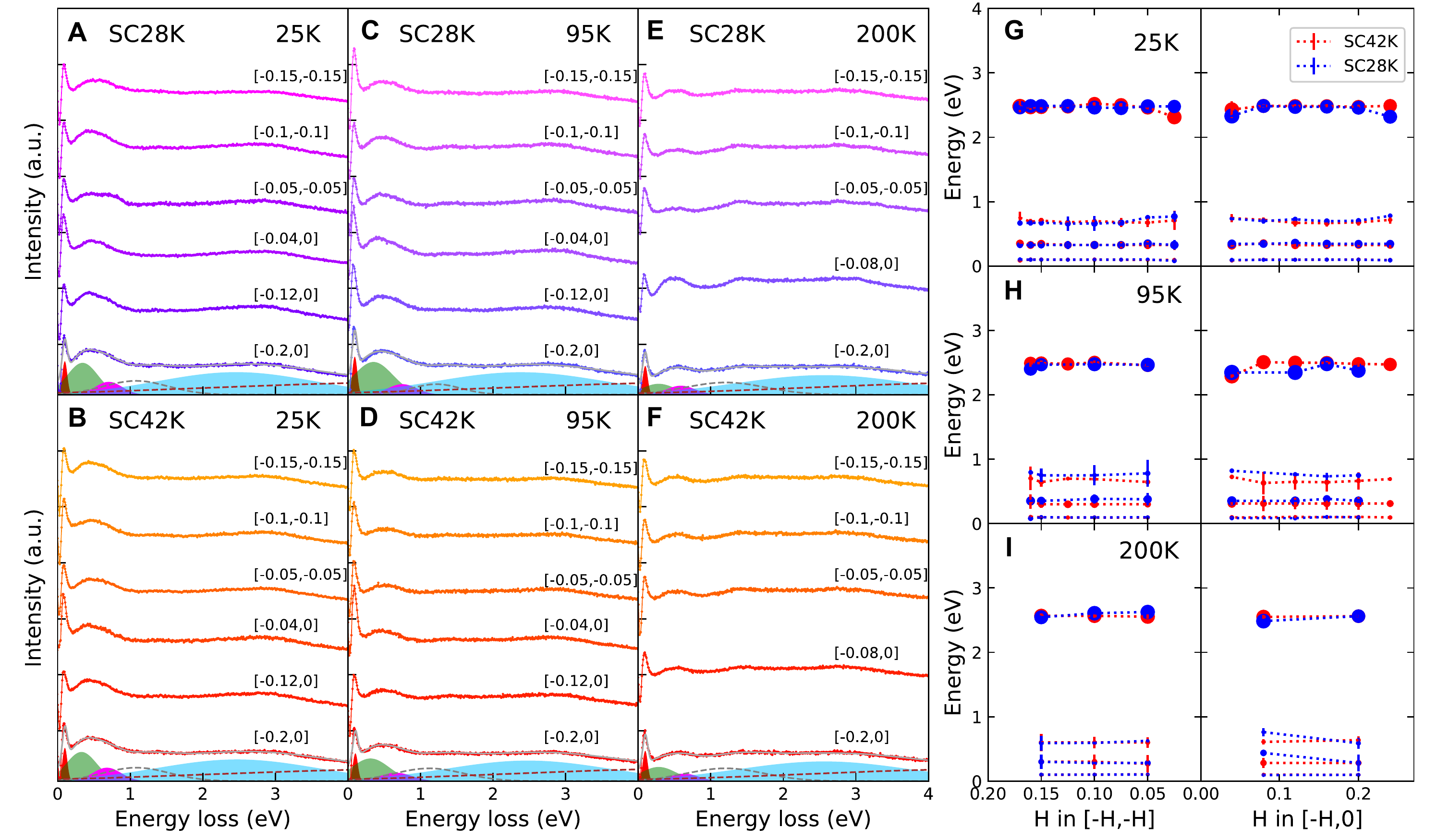}
\caption{{\bf Momentum-dependent RIXS spectra of FeSe11111 at 25\,K, 95\,K and 200\,K}. ({\bf A}), ({\bf C}), ({\bf E}) Selected RIXS spectra of SC28K along two high-symmetry directions as indicated in Fig.~\ref{fig1}({C}) at 20\,K, 95\,K and 200\,K, respectively. Each spectrum is shifted
vertically for clarity. The spectra were collected at the Fe-$L_3$ resonant peak energy of 709\,eV. Spectra were normalized to the high energy-loss range of [4, 12] eV. The elastic peak was subtracted from data to visualize the low energy features. ({\bf B}), ({\bf D}), ({\bf F}) Similar measurements on SC42K. Examples of the six peak decompositions described in text are shown in the bottom spectra. ({\bf G}), ({\bf H}), ({\bf I}) Summary of fitted energies of four Raman features along two high-symmetry directions for both SC28K and SC42K at 25\,K, 95\,K and 200\,K, respectively. The size of the markers corresponds to the spectral weight of the features. Error bars were the standard deviation of the fits.
\label{fig2}}
\end{figure*}

\begin{figure*}[htbp]\hspace{-5mm}
\centering
\includegraphics[width=\linewidth]{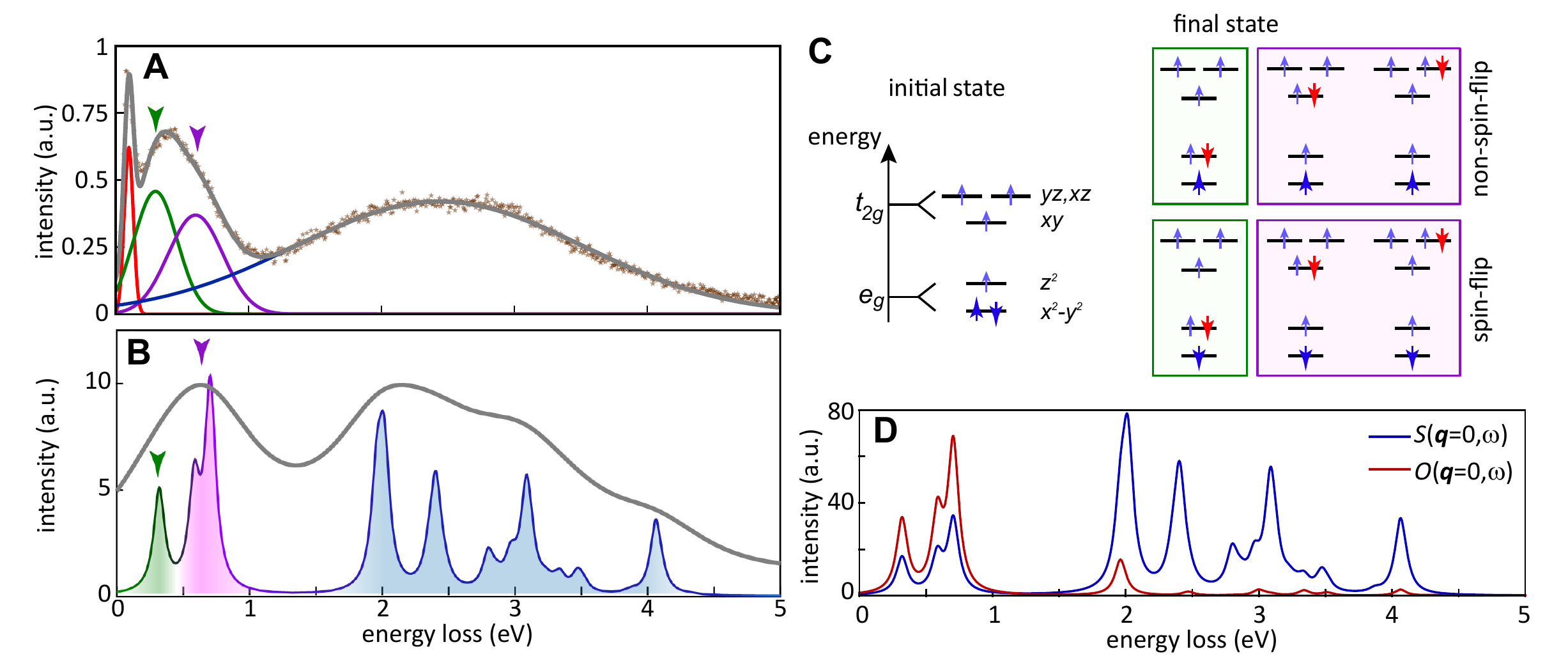}
\caption{{\bf Experimental and theoretical RIXS spectra.}  ({\bf A}) The same RIXS spectrum as that in Fig.~\ref{fig1}(D) while the elastic peak was subtracted from data, leaving only four Raman features. 
The colored solid lines are fitting results of each Raman feature. The gray solid line is the sum of the fitting results of the four Raman peaks.
({\bf B}) RIXS spectra simulated by atomic multiplet theory. The broad peak at $\sim 0.5$\,eV consists of 0.3\,eV and 0.7\,eV excitations, which are respectively $d-d$ transitions among the $e_g-e_g$ and $e_g - t_{2g}$ orbitals.
The broad 2.5\,eV high-energy peak consists of multiple excited states of total spin quantum number $S=1$.
({\bf C}) Schematic cartoons illustrating the spin and orbital characters of the ground state and the low-energy excited states. ({\bf D}) Dynamical spin structure factor $S(\mathbf{q}=0,\omega)$ and (non-spin-flip) orbital structure factor $O(\mathbf{q}=0,\omega)$ calculated using the multiplet model defined in the text.
\label{fig3}}
\end{figure*}

\noindent
{\bf {RESULTS}}\\
\noindent
{\bf Excitations in FeSe11111}\\
Figure~\ref{fig1}A shows the crystal structure of the FeSe-based superconductor FeSe11111 in the one-Fe unit notation. It consists of anti-PbO-type FeSe layers sandwiched by anti-PbO-type (Li/Fe)OH layers along c-direction, and belongs to the space group P4/nmm (No. 129)\,\cite{lu2015}. The schematics of the RIXS experimental geometry is shown in Fig.~\ref{fig1}B. 
The accessible reciprocal space at Fe $L_3$-edge is displayed in Fig.~\ref{fig1}C. We state coordinates in units of 2$\pi$/a, i.e., 1 corresponds to one reciprocal lattice unit (r.l.u.).  

Figure~\ref{fig1}D shows a representative RIXS spectrum of FeSe11111 collected at the Fe $L_3$-edge resonant energy (709\,eV) using $\pi$-polarized incident x-rays. 
Intriguingly, the RIXS spectrum of FeSe11111 highly resembles that of monolayer FeSe, including the weak fluorescence background and broad features at around 0.5\,eV and 3\,eV\,\cite{Jonathan2021.natcom}. This indicates that these features arise from the FeSe layer instead of the Li-Fe layer. Moreover, we obtained much better statistics of RIXS spectra in FeSe11111 than those in monolayer FeSe due to the larger sample volume. It is noteworthy that, due to the metallic nature of Fe-based superconductors, the fluorescence background is substantially strong in the RIXS spectra of bulk FeSe\,\cite{M.C.Rahn.prb.FeSe,Jonathan2021.natcom} and iron pnictides\,\cite{yang2009prb,BFA.RIXS.natcom}. Here, the weak fluorescence background is probably due to the absence of hole Fermi-surface pockets at $\Gamma$ point, which closes the particle-hole scattering channel and allows us to identify clearly the excitations in FeSe11111.
To quantitatively analyze the data, we fitted the RIXS spectra with Gaussian profiles and identified six excitations as shown in Fig.~\ref{fig1}D, which are located at $\sim$ 0.1\,eV, $\sim$ 0.3\,eV, $\sim$ 0.7\,eV, $\sim$ 1.1\,eV, $\sim$ 2.5\,eV and $\sim$ 4.8\,eV, respectively [see supplementary Note S4 for fitting details]. 

Since the incident-energy detuning RIXS measurements can distinguish between localized excitations and delocalized excitations involving continua\,\cite{zhoukj.detuning,bisogni2016ground}, we measured a series of RIXS spectra across Fe $L_3$-edge in steps of 0.25\,eV for SC42K (Fig.~\ref{fig1}E). 
For localized excitations, the emitted photon energy varies by the same amount as the variation of incident photon energy, which is referred as Raman-like features. While for delocalized excitations, the energy loss varies by the same amount as the variation of incident photon energy, which is referred as fluorescence-like features\,\cite{RevModPhys.RIXS.2011}. 
The Raman-like features are identified at around 0.1\,eV, 0.3 $\sim$ 0.7\,eV and 2.5\,eV, which will be our focus in the following. The features at around 1.1 eV and 4.8 eV energy loss at the Fe $L_3$-edge resonant energy tracks the incident energy [see Fig.~\ref{fig1}E and F], which are typical fluorescence behaviors. They correspond to the Fe 3d to 2p fluorescence decay reported also in iron pnictides and Fe 3d to Se 4p fluorescence decay, respectively\,\cite{yang2009prb,A.Subedi2008prb,Liu_2017}. We obtained similar detuning results in SC28K (see Supplementary Fig. S4).\\

\begin{figure*}[htbp]
\centering
\includegraphics[width=\linewidth]{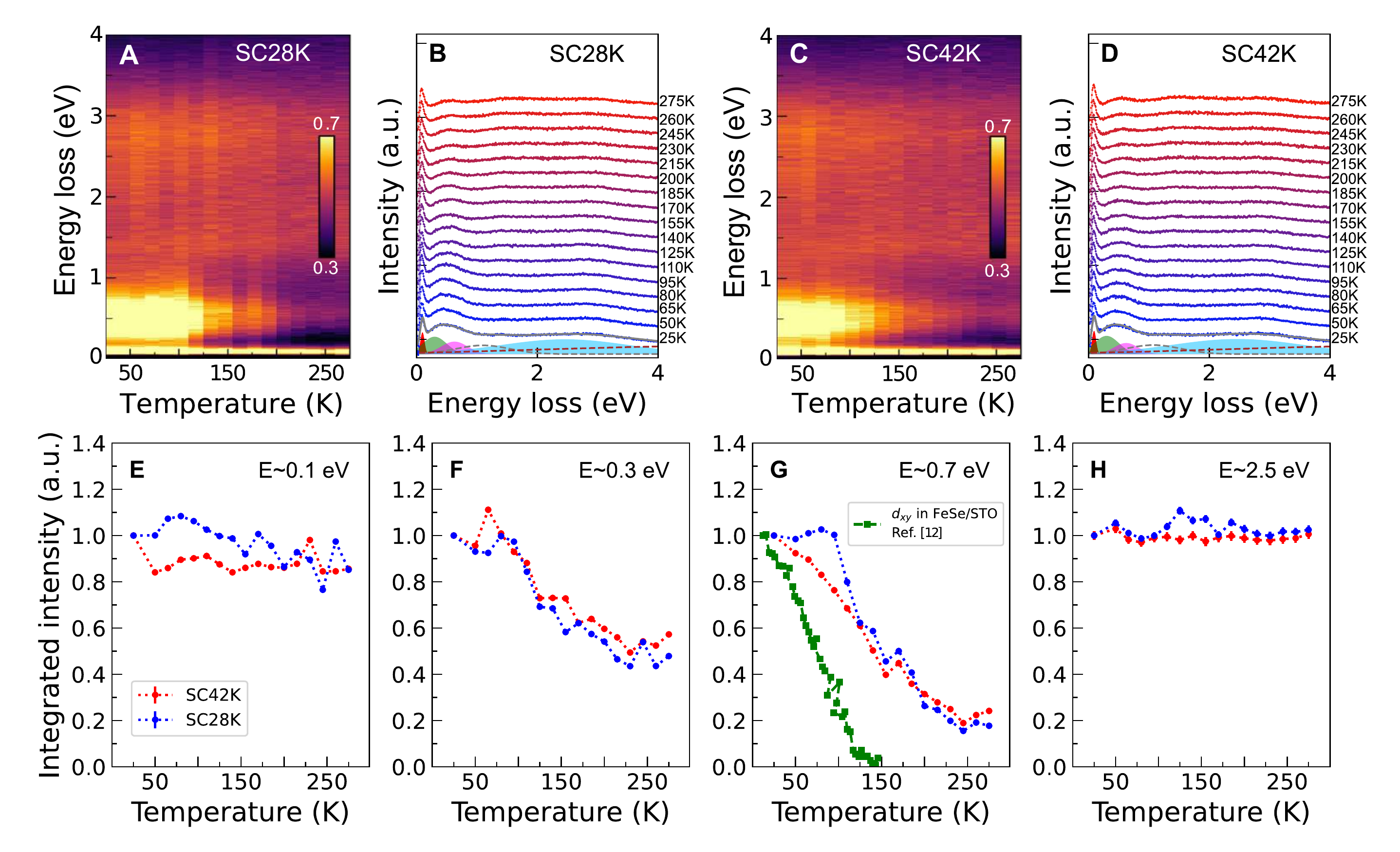}
\caption{{\bf Temperature dependence of SC28K and SC42K at {\bf q}= [-0.2, 0]}. Spectra were normalized to the integrated fluorescence intensity from -4 eV to -12 eV. ({\bf A}), ({\bf C}) Energy vs temperature RIXS maps measured on SC28K and SC42K, respectively. The elastic peak was subtracted to manifest the low-energy excitations. ({\bf B}), ({\bf D}) The waterfall plots correspond to ({\bf A}) and ({\bf C}), respectively. Examples of the fitting components are shown in the bottom spectra. ({\bf E})-({\bf H}) Temperature dependence of the fitted integrated intensity of these four Raman features, respectively. The integrated intensity was normalized to that of 25 K for comparison. Temperature evolution of spectral weight of $d_{xy}$ orbital measured by ARPES in monolayer FeSe/STO (green markers) is shown in ({\bf G}) for comparison\,\cite{ARPES.2015.MYi}. 
\label{fig4}}
\end{figure*}

\noindent
{\bf Momentum and $T_{\mathrm{c}}$ dependence}\\
To understand the Raman-like excitations we measured the detailed momentum-resolved RIXS maps along two high symmetry directions, [0, 0]$-$[-H, 0] and [0, 0]$-$[-H, -H], as shown in Fig.~\ref{fig2}. 
The elastic peaks were subtracted to manifest the low-energy excitations. The fittings of RIXS spectra are shown in the bottom spectra in Fig.~\ref{fig2}A-F and also in the Supplementary Fig. S6. The Raman-like excitations at $\sim$ 0.1 eV, 0.3 eV, 0.7 eV and 2.5 eV barely change with momenta along the two high symmetry directions. 
The two excitations at $\sim$ 0.3 eV and 0.7 eV have similar half width at half maximum (HWHM) between 0.2 and 0.3 eV (see Supplementary Fig. S7), which is much broader than our energy resolution ($\sim$ 0.08 eV). They resemble those around 0.5 eV in monolayer FeSe, which were attributed to spin excitations\,\cite{Jonathan2021.natcom}. 
These excitations have similar energies in both SC28K and SC42K, irrespective of varying $T_{\mathrm{c}}$ (Fig.~\ref{fig2}G). We also measured the momentum-dependent RIXS spectra at 95\,K and 200\,K (Fig.~\ref{fig2}C-F), which showed the persistence of these Raman-like excitations to high temperatures. While their energies do not change with temperature (Fig.~\ref{fig2}G-I), the intensities descend with increasing temperature (Fig.~\ref{fig2}A-F). We will discuss the temperature dependence in more detail below.

The dispersionless behavior of the Raman-like features suggests that the spectral peaks should be of local nature. Different from $d^9$ copper oxide supercondutors where the low-energy states lie in a single $3d$ orbital, the hybridization and interactions among five partially filled $3d$ orbitals in Fe allow rich atomic excitations, which can be reflected in the RIXS spectra. Using atomic multiplet theory with a distorted tetrahedral crystal field environment (CFE), we were able to reproduce theoretical spectral peaks at the experimentally observed excitation energies [see comparisons in Fig.~\ref{fig3}A and B].
By projecting the excited-state wavefunctions to the maximal-overlapped Fock states, the nature of each excitation can be addressed approximately in a single-particle picture [see Fig.~\ref{fig3}C]. The $\sim 0.3$ and $0.7$\,eV features, which are more evident in experiments, correspond to $d-d$ excitations between the $e_g-e_g$ and $e_g-t_{2g}$ orbitals, respectively. These excitations contain both spin-flip and non-spin-flip contributions, which can be further distinguished through the polarization control (See Fig.~\ref{fig3}D and SI).
Due to the locality of these low-energy $d-d$ excitations, their energy scales are primarily determined by crystal field and lattice geometry, instead of in-plane hoppings or Coulomb interactions. Therefore, these excitations are robust against varying $T_{\mathrm{c}}$. 
We notice that RIXS measurements on FeS complexes with a distorted tetrahedral CFE for ${\mathrm{Fe}}^{2+}$ also observed two peaks at 0.32 and 0.58 eV attributed to $d-d$ excitations\,\cite{2018.RIXS.IronSulfur}. RIXS experiments on PrFeAsO$_{0.7}$ ($T_{\mathrm{c}}$=42\,K) revealed a $\sim$ 0.5\,eV feature assigned to $d-d$ orbital excitation as well\,\cite{PhysRevB_PrFeAsO0.7}.

In addition to these two low-energy peaks, our simulation exhibits substantial intensity in a 2 eV window around the energy loss $\omega\sim2.5$ eV, which consists of multiple spectral peaks when the spectral broadening is reduced. These excited states mainly exhibit a change of total spin quantum number $\Delta S =1$. 
The energy scale for these high-energy excitations is mainly determined by the Racah parameters $B$ and $C$ between $d-d$ orbitals. In our calculation, we used $B = 0.12$ eV and $C=0.4$ eV, which lead to a typical Hund's coupling $\sim 0.8-0.9$ eV for Fe$^{2+}$.
When only a single Fe site is present, the Racah parameter $A$ provides only an overall energy shift of the atomic multiplet levels, and it is simply set to zero here.
Since $B$ and $C$ arise from multipole interactions, they are more difficult to be screened (unlike $A$ due to monopole interaction). Therefore, it is expected that the high-energy 2.5 eV feature will not be strongly affected by charge screening and the change of $T_{\mathrm{c}}$, which is also consistent with the experiment.\\

\noindent
{\bf Temperature dependence}\\
To further explore the nature of the Raman excitations, we performed temperature dependent measurements from 25\,K to 275\,K on both SC28K and SC42K (Fig.~\ref{fig4}). It is clear that the intensity at the energy-loss region of 0.2 $\sim$ 0.8\,eV decreases with temperature (Fig.~\ref{fig4}A,C). Moreover, the ratio of the intensity between two excitations at $\sim$ 0.3\,eV and $\sim$ 0.7\,eV varies with temperature (Fig.~\ref{fig4}B,D), which supports our fitting with two peaks (details of fitting shown in Supplementary Fig. S8). The temperature dependence of the integrated intensity of the four Raman features is shown in Fig.~\ref{fig4}E-H. Intriguingly, the excitations at $\sim$ 0.3\,eV and $\sim$ 0.7\,eV decrease by more than 50$\%$ from 25\,K to 275\,K, while the features at $\sim$ 0.1\,eV and $\sim$ 2.5\,eV change only slightly with temperature in both samples. It is also known that the change of $T_{\mathrm{c}}$ is connected to different Fe vacancy concentrations\,\cite{FeSe11111Tc42K,Yiyuan.CPL,FeVacancy2019prm}. This can affect the oxidation state, so that Fe valency may be reduced from $2+$, which can slightly affect the ratio between the 0.3\,eV and 0.7\,eV peaks.

The temperature dependence of the 0.3 eV and 0.7 eV spectral features suggests the primary orbital nature of these excitations. Previous magnetic susceptibility measurements on SC28K and SC42K displayed a drop or deviation from the Curie-Weiss behavior around $\sim$ 120K, implying a magnetic transition or fluctuation \,\cite{Yiyuan.CPL,Dong.prb}. In Fe-based superconductors, magnetic transitions or fluctuations are usually accompanied by an orbital weight redistribution. Previous studies have demonstrated that the spectral intensity can exhibit a more apparent temperature dependence when orbital fluctuations are present\,\cite{Chen2009}. To demonstrate the relevance of orbital fluctuations, we simulate the dynamical spin and (non-spin-flip) orbital structure factors $S(\mathbf{q}=0,\omega)$ and $O(\mathbf{q}=0,\omega)$ separately (see Fig.~\ref{fig3}D). Orbital excitations with spin-flip character is included in the $S(\mathbf{q},\omega)$. As expected from the single-particle picture in Fig.~\ref{fig3}C, the two low-energy Raman peaks at $\sim$ 0.3 eV and $\sim$ 0.7 eV are present in both orbital and spin channels. Considering the experimental measurements integrated over all scattering polarizations, the contributions from these two channels are approximated 2:1. Since the onset of orbital excitations can cause a rapid transfer of the spectral weight away from coherent peaks\,\cite{Chen2009}, a rough estimation is that the intensity of these $d-d$ excitations drops by 60\% at high temperature due to the fluctuation of orbitals, consistent with the experimental observations for SC28K and SC42K. Moreover, we notice the suppression of the $\sim$ 0.7 eV peak involving the $d_{xy}$ orbital is consistent with ARPES measurements on monolayer FeSe that show the disappearance of $d_{xy}$ spectral weight at high temperature\,\cite{ARPES.2015.MYi} [see Fig.~\ref{fig4}G]. This agreement implies substantial orbital fluctuations and orbital redistribution in high $T_c$ iron chalcogenides.

In contrast, the orbital $O(\mathbf{q}=0,\omega)$ has almost no contribution at the high-energy regime. Thus, the 2.5 eV Raman feature originates primarily from local spin-flip $d-d$ excitations, accounting for its weak temperature dependence.
This mechanism can be further tested in the future by the temperature dependence of RIXS spectra employing a polarimeter for full polarization measurements. In addition to the above three features at higher energies, another Raman feature at $\sim$ 0.1 eV is also dispersionless but is absent in the multiplet model calculation. 
ARPES and electron energy-loss spectroscopy (EELS) measurements on monolayer FeSe reported a phonon mode at $\sim$ 90-100 meV arising from the STO substrate\,\cite{2017.prl.monolyerFeSe.ARPES.,2014.nat.monolayerFeSe.ARPES,shuyuanzhang.prb.2016}. However, our FeSe11111 thin-film has a different substrate ${\mathrm{LaAlO}}_{3}$ and the thickness of the film is of several hundreds of nanometers. A previous Raman study on FeSe11111 measured phonons only up to 400 $cm^{-1}$\,\cite{Raman.FeSe11111}. Nevertheless, the presence of oxygen and hydrogen within (Li/Fe)OH layers may contribute to a 0.1 eV phonon excitation. Another possibility relates to an optical spin wave mode, which is typically much less dispersive, compared to the acoustic spin wave mode. In this case, the system exhibits short- or long-range magnetism involving block spins in an enlarged magnetic unit cell\,\cite{INS.RbFeSe}. However, INS studies of FeSe11111 have not reported an optical magnon around 0.1 eV\,\cite{BingyingPan.FeSe1111.ins2017natcom}, which renders this origin less likely. \\

\noindent
{\bf {DISCUSSION}}\\

The dispersionless excitations in FeSe11111 and monolayer FeSe contrast the dispersive spin excitations observed in bulk FeSe, reflecting their intrinsic differences. Our theory-experiment comparison suggests that a minimal multiplet model involving the Fe 3$d$ orbitals can quantitatively address these dispersionless features. Informed by experimental results, our simulation ignores nonlocal interactions which usually cause dispersive magnon excitations. That said, an orbital-dominated mechanism does not contradict and may coexist with the recently proposed scenario of hybridized magnons in a two-band Hubbard model\,\cite{Jonathan2021.natcom}. In a wider momentum range, our RIXS results are compatible with the INS in FeSe11111, which revealed that spin excitations near [$\pi$,$\pi$] persist to 140 meV\,\cite{BingyingPan.FeSe1111.ins2017natcom}. 
Moreover, due to the effect of electron-doping, the spectral weight of the high-energy spin excitations in FeSe11111 is much lower than that in bulk FeSe \,\cite{BingyingPan.FeSe1111.ins2017natcom,QisiWang.FeSe.ins2016natcom}, which would make it difficult to be observed by RIXS. 

Orbital-selective Cooper pairing has been discovered in bulk FeSe using Bogoliubov quasiparticle interference imaging \cite{FeSe_Orbitalpairing}. Here, the general existence of these distinct orbital excitations in both FeSe11111 and monolayer FeSe\,\cite{Jonathan2021.natcom} suggests that orbital fluctuations contribute for the significant enhancement of the superconducting transition temperature.
In FeSe11111 without any substrates, these orbital fluctuations have shown sensitivity to the lattice geometry. Therefore, these fluctuations can be further enhanced when growing a monolayer FeSe on the STO substrate, correlating with the enhancement of $T_c$. This mechanism is consistent with the previous scanning tunneling microscopy (STM) studies in both monolayer FeSe and FeSe11111\,\cite{prbFeSe11111plain,natp.monolayerFeSe.plain}, in favor of the phonon-mediated pairing enhancement scenario\,\cite{2014.nat.monolayerFeSe.ARPES} or orbital-fluctuation-mediated pairing mechanism\,\cite{PhysRevLett_orbitalfluctuation}. The absence of the $d_{xz}/d_{yz}$ hole Fermi surface may also suppress the nematic order that competes with superconductivity\,\cite{2017ShiX.natc.monoFeSe.ARPES}, underscoring the crucial role of the orbitals in the pairing. Thus, as a layered material without STO substrate, FeSe11111 and its excitations provide an unique opportunity to elucidate the evolution from bulk to monolayer FeSe materials, which exhibits an substantial increase of $T_c$. \\

\noindent
{\bf {METHODS}}\\
{\bf Sample preparation}\\
The high-quality single-crystalline FeSe11111 films were synthesized by the matrix-assisted hydrothermal epitaxy (MHE) technique\,\cite{sample.method1, huang2017arxiv}. Their crystalline quality were characterized by x-ray diffraction (XRD) on a 9 kW Rigaku SmartLab x-ray diffractometer equipped with two Ge(220) monochromators at room temperature (Supplementary Fig. S1). Both the XRD patterns exhibit a single preferred (00l) orientation and the full width at half maximum (FWHM) is around $0.3^\circ$ of the x-ray rocking curves for the (006) reflection, showing the highly crystalline quality of these films (Supplementary Fig. S1). Their $T_{\mathrm{c}}$ values are defined at the onset temperatures of the diamagnetism by magnetic measurements (Supplementary Fig. S1), performed with a 1 Oe magnetic field applied along the c-axis using a SQUID magnetometer (MPMS XL1, Quantum Design). \\

\noindent
{\bf RIXS measurements}\\
We performed high-resolution RIXS experiments by using the Super Advanced x-ray Spectrometer (SAXES) at the Advanced Resonant Spectroscopy (ADRESS) beamline of the Swiss Light Source, Paul Scherrer Institut, Switzerland\,\cite{psi1,psi2}. The energy resolution was 80 meV (FWHM). We cleaved both samples in ultra-high vacuum. We used the grazing-incident geometry which provided stronger signals than the grazing-out geometry (Supplementary Fig. S3). All samples were aligned with the surface normal (001) in the scattering plane. $\pi$ polarized x-rays were used for the RIXS measurements since the signals were similar between $\pi$- and $\sigma$-polarizations (Supplementary Fig. S3). The incident energy was tuned to the peak maximum of Fe $L_3$-edge (709 eV), except for detuning measurements. The angle between incoming and outgoing x-rays was fixed to 130 degrees, resulting in a constant value of the total momentum transfer. We refer to reciprocal space coordinates in the conventional one-Fe unit cell with $a$ = $b$ = 2.680 \AA, $c$ = 9.318 {\AA} for SC42K and $c$ = 9.261 {\AA} for SC28K\,\cite{Yiyuan.CPL}. The momentum transfer {\bf Q} is defined as {\bf Q} = {\bf H}$a^*$ + {\bf K}$b^*$ + {\bf L}$c^*$ where $a^*$ = 2$\pi$/$a$, $b^*$= 2$\pi$/$b$, and $c^*$ = 2$\pi$/$c$. {\bf q} is the projection of {\bf Q} in ab-plane. We measured ab-plane momentum transfer scans by varying the grazing incident angle to the sample surface. Each RIXS spectrum was measured for 30 min except for the spectrum of incident energy dependent RIXS maps, which was measured for 5 min. \\

\noindent
{\bf Atomic Multiplet Theory}\\
To model the experimental spectra, we make two assumptions about the Hamiltonian and the RIXS cross section. First, we consider only multiplet interactions in the atomic limit\,\cite{de2005multiplet}. In this case, our calculation can capture local spin or orbital flip excitations, but not collective spin waves (magnon). Nevertheless, this assumption is appropriate in our system, as (i) the experiment exhibits little momentum dependence, and (ii) we are not aware of any lattice model with short or long range magnetism capable of exhibiting a complete flat spin-wave dispersion. The resulting multiplet Hamiltonian containing all Fe $3d$ orbitals reads
\begin{eqnarray}\label{eq:Hubbard}
	\mathcal{H}_{3d} &=& \sum_{\alpha} U_{\alpha} d_{\uparrow}^{(\alpha)\dagger}d_{\uparrow}^{(\alpha)} d_{\downarrow}^{(\alpha)\dagger}d_{\downarrow}^{(\alpha)}\nonumber\\
	& &+ \frac12 \sum_{\alpha\neq\beta\atop \sigma\sigma^\prime}U^{\prime}_{\alpha\beta} d_{\sigma}^{(\alpha)\dagger}d_{\sigma}^{(\alpha)} d_{\sigma^\prime}^{(\beta)\dagger}d_{\sigma^\prime}^{(\beta)} + \sum_{\alpha, \sigma} \varepsilon_{\alpha} d_{\sigma}^{(\alpha)\dagger}d_{\sigma}^{(\alpha)}\nonumber\\
	& &+ \frac12 \sum_{\alpha\neq\beta\atop \sigma\sigma^\prime}J_{\alpha\beta} d_{\sigma^\prime}^{(\alpha)\dagger}d_{\sigma}^{(\beta)\dagger} d_{\sigma}^{(\alpha)}d_{\sigma^\prime}^{(\beta)}.
\end{eqnarray}
Here, $d_{\sigma}^{(\alpha)}$ ($d_{\sigma}^{(\alpha)\dagger}$) annihilates (creates) a $3d_{\alpha}$ electron with spin $\sigma$ ($\alpha=x^2-y^2$, $z^2$, $xy$, $yz$, and $xz$). $\varepsilon_{\alpha}$ denotes the site energy of orbital $\alpha$. In a distorted tetragonal CFE, $\epsilon_\alpha$ can be parameterized by three parameters 10$Dq$, $\delta$, and $\mu$\,\cite{chilkuri2017revisiting}. We use 10$Dq=0.485$ eV, $\delta = -0.12$ eV, and $\mu=-0.32$ eV. Here, 10$Dq$ represents the site energy separation between $e_g$ and $t_{2g}$ orbitals, and its typical value is $\sim 0.5$ eV for a tetrahedral CFE in iron-based superconductors. $\delta$ and $\mu$ describe additional splitting between the $e_g$ and $t_{2g}$ orbitals due to a lowered crystal symmetry.
The intra-orbital $U$ and inter-orbital $U^{\prime}$ interactions, as well as Hund's coupling and pair hopping $J$, can be parameterized by the Racah parameters $A$, $B$, and $C$. In the case of a single Fe atom, $A$ only provides an overall energy shift, so it is simply set to zero. Without any fine tuning, we use $B$=0.12 eV and $C$=0.40 eV, leading to a typical value of $\sim$ 0.8 -0.9 eV for Fe$^{2+}$ Hund's coupling, independent of charge screening and the change of $T_{\mathrm{c}}$. The Hamiltonian is solved by exact diagonalization (ED) to obtain the energy eigenvalues and eigenstates.
It is noted that within a reasonable CFE and interaction parameters, a high spin $S=2$ ground state is always favored, which agrees with our further {\it ab initio} CASPT2 simulations (see below). If the ground state spin quantum number is reduced from $S=2$, then the spectral intensity ratio between the 0.3 eV and 0.7 eV peaks also will change\,\cite{thole1988branching}. This can happen with charge or orbital fluctuation effects, which are always present to some extent in the experiment but not captured in the atomic multiplet calculation.

The RIXS cross section is defined as
\begin{equation}
	I(\mathbf{q},\omega,\omega_{\rm in}) = \frac1{\pi N} \mathrm{Im}\langle \Psi | \frac1{\mathcal{H}_{3d} - e - \omega - i\delta}  |\Psi\rangle,
\end{equation}
where
\begin{equation}
	|\Psi\rangle = \frac1N\sum_{\mathbf{k}}\langle \mathcal{D}_{\mathbf{k}+\mathbf{q}}^\dagger \frac1{\mathcal{H}_{3d} + H_{\textrm{core}}- e - \omega_{\rm in} - i\Gamma} \mathcal{D}_{\mathbf{k}} |G\rangle,
\end{equation}
and the (momentum-resolved) dipole transition operator $\mathcal{D}_{\mathbf{k}} = \sum_\mathbf{i} e^{-\mathbf{k}\cdot \mathbf{r}_\mathbf{i}} \mathcal{D}_{\mathbf{i}}$.
In principle, the intermediate state of a RIXS process contains a core-hole interaction (with core-hole lifetime $1/\Gamma$) $H_{\textrm{core}}$, describing the interaction between Fe core $2p$ and valence $3d$ orbitals. 
To simplify the calculation and clearly assign the spectral features, we adopt the second commonly used assumption of ultrashort core-hole lifetime approximation\,\cite{van2005epl, ament2007prb}, i.e.~$\Gamma\gg U_\alpha$. In this case, the RIXS cross section is reduced to the dynamical spin and orbital structure factors. The dynamical spin structure is defined as
\begin{equation}
    S(\mathbf{q},\omega) = \frac{1}{\omega N}\textrm{Im}\left\langle G\right|\rho_{-\mathbf{q}}^{(s)}\frac{1}{H_{3d}-e-\omega-i\delta}\rho_\mathbf{q}^{(s)}\left|G\right\rangle,
\end{equation}
where $\rho_\mathbf{q}^{(s)} = \sum_{\alpha\beta} d_{\mathbf{k}+\mathbf{q}\uparrow}^{(\beta)\dagger} d^{(\alpha)}_{\mathbf{k} \downarrow}$ and the orbital excitation is defined as
\begin{equation}
   O(\mathbf{q},\omega) = \frac{1}{\omega N}\textrm{Im}\left\langle G\right|\rho_{-\mathbf{q}}^{(o)}\frac{1}{H_{3d}-e-\omega-i\delta}\rho_\mathbf{q}^{(o)}\left|G\right\rangle,
\end{equation}
where $\rho_\mathbf{q}^{(o)} = \sum_{\alpha\neq\beta\sigma} d_{\mathbf{k}+\mathbf{q}\sigma}^{(\beta)\dagger} d^{(\alpha)}_{\mathbf{k} \sigma}$. These two structure factors approximate the spin-flip and non-spin-flip RIXS at the ultrashort core-hole lifetime limit.
We note that the charge structure factor $N(\mathbf{q},\omega)$ also contributes to the non-spin-flip cross section, but its inelastic response vanishes for $\mathbf{q}=0$ due to charge conservation. Using the atomic multiplet model, we only present the response function for $\mathbf{q}=0$ in this paper.\\

\noindent
{\bf Ground-State Configuration from \emph{ab initio} Quantum Chemistry}\\

\begin{table}[]
    \centering
    \begin{tabular}{c|cc}\hline\hline
      spin configuration & DFT (-295.7 keV) & CASPT2 (-298.86 keV) \\ \hline
        $S=0$ & -1.981187\,eV & -4.161511\,eV\\
    $S=1$  & -3.609398\,eV & -4.271853\,eV\\
    $S=2$  & -5.065212\,eV &  -5.049349\,eV\\\hline\hline
    \end{tabular}
    \caption{Ground-state energy for each spin configuration calculated using DFT and CASPT2.}
    \label{tab:qcGroundState}
\end{table}

To determine the ground-state spin configuration as a starting-point for the ED calculation, we perform two different \emph{ab initio} simulations for a single unit cell Fe(II)[Se$^{2-}]_4$ complex. Due to the expectation of strong multi-reference effects, we complement the PBE0 DFT simulation with a CASPT2 simulation with (10e,12o) active space. As shown in Table~\ref{tab:qcGroundState}, both simulations conclude with a ground state with $S=2$, justifying the selection of CFE in our atomic multiplet simulations. More specifically, the DFT simulations predict relatively large energy differences among all spin sectors, compared to the CASPT2 results. This is because the lower-spin state exhibits a stronger multi-reference effect as a result of quantum fluctuations. Without accounting for this effect, DFT over-estimated the energy of these lower-spin states. In contrast, the energy differences obtained by CASPT2 are relatively small, accounting for the experimental observation of reduced magnetic moment with the presence of itineracy.

For DFT calculation, we used the PBE0 hybrid function (PBE GGA + 20\% Hartree-Fock exchange) with def2-TZVP basis set, which has been frequently used in spin splitting energy calculations of transition metal complexes\,\cite{Gani, Liu}. The simulation is conducted using ORCA 4.0\cite{neese2020orca}.
For CASPT2 calculation, we followed literature recommendations for  transition metals to include five $3d$ orbitals\,\cite{pierloot2003,veryazov2011select}, two $\sigma$ orbitals describing metal-ligand bonding, and five double-shell $d$ orbitals for mid-row and later transition, resulting in 12 active orbitals.  We employ relativistic atomic natural orbital (ANO-rcc) basis sets contracted to [7s6p5d3f2g1h] for Fe and [7s6p4d3f2g] for Se\,\cite{Douglas1974, Hess1986}. The 41 core orbitals were frozen in all calculations. The CASPT2 simulation is conducted using Molcas 8\cite{aquilante2016molcas}.\\

\noindent
{\bf DATA AVAILABILITY}\\
All data needed to evaluate the conclusions in the paper are present in the paper and/or the Supplementary Materials. Additional data related to this paper may be requested from the authors.

\bibliographystyle{naturemag}
\bibliography{reference}

\vspace{1 ex}
\noindent
{\bf ACKNOWLEDGMENTS}\\ 
This work was performed at the ADRESS beamline of the Swiss Light Source using the
SAXES instrument jointly built by Paul Scherrer Institut, Switzerland, Politecnico di
Milano, Italy and EPFL, Switzerland. 
The calculations were performed at the Texas Advanced Computing Center using the Frontera supercomputer funded by the National Science Foundation.
We appreciate the help with sample preparation from Yong Hu, Aiji Liang, Li Yu and Chennan Wang. We acknowledge valuable discussion with Yuan Li, Yan Zhang, Fa Wang, Hlynur Gretarsson, Jonathan Pelliciari and Steve Johnston. \\

\noindent
{\bf AUTHOR CONTRIBUTIONS}\\ 
Y.Y.P conceived and designed the experiments with suggestions from T.S.; Q.X., W.L.Z., T.C.A., Y.T., T.S. and Y.Y.P. performed the RIXS experiment at the Swiss Light Source with the help of Q.Z.L. and S.L.Z.. D.L. and X.L.D. synthesized, grew and characterized the FeSe11111 thin-films; Y.Y.P. and Q.X. analysed the RIXS experimental data; C.C.C. and Y.W. performed the calculations; Y.Y.P., Q.X., C.C.C., and Y.W. wrote the manuscript with input and discussion from all co-authors.\\

\noindent
{\bf FUNDING}\\
Y.Y.P. is grateful for financial support from the Ministry of Science and Technology of China (Grant No. 2019YFA0308401) and the National Natural Science Foundation of China (Grant No. 11974029). T.S. acknowledges support by the Swiss National Science Foundation through
Grant Numbers 200021\_178867, CRSII2\_160765/1 and CRSII2\_141962. T.C.A. acknowledges funding from the European Union’s Horizon 2020 research and innovation programme under the Marie Skłodowska-Curie grant agreement No. 701647 (PSI-FELLOW-II-3i program). X.L.D. is grateful for financial support from the National Natural Science Foundation of China (Grant Nos. 12061131005, 11834016 and 11888101), the Strategic Priority Research Program of Chinese Academy of Sciences (XDB25000000). \\ 

\noindent
{\bf COMPETING INTERESTS}\\
The authors declare that they have no competing interests.\\

\end{document}